\documentclass[twocolumn,showpacs,amsmath,amssymb]{revtex4}
\usepackage{graphicx}
\usepackage{color}
\begin{document}

\title{Power-law Ansatz in Complex Systems: Excessive Loss of Information}
\author{Sun-Ting Tsai$^{1}$, Chin-De Chang$^{1}$, Ching-Hao Chang$^{1}$, Meng-Xue Tsai$^{1}$, Nan-Jung Hsu$^{2}$, and Tzay-Ming Hong$^{1,\star}$}
\affiliation{\\$^1$Department of Physics, National Tsing Hua University, Hsinchu 30013, Taiwan, Republic of China
\\$^2$Institute of Statistics, National Tsing Hua University, Hsinchu 30013, Taiwan, Republic of China}
\date{\today}
\begin{abstract}
The ubiquity of power-law relations in empirical data displays physicists' love of simple laws and uncovering common causes among seemingly unrelated phenomena. However, many reported power laws lack statistical support and mechanistic backings, not to mention discrepancies with real data are often explained away as corrections due to finite size or other variables. We propose a simple experiment and rigorous statistical procedures to look into these issues. Making use of the fact that the occurrence rate and pulse intensity of crumple sound obey power law with an exponent that varies with material, we simulate a complex system with two driving mechanisms by crumpling two different sheets together.  The probability function of  crumple sound is found to transit from two power-law terms to a {\it bona fide} power law as compaction increases. In addition to showing the vicinity of these two distributions in the phase space, this observation nicely demonstrates the effect of interactions to bring about a subtle change in macroscopic behavior and more information may be retrieved if the data are subject to sorting. Our analyses are based on the Akaike information criterion that is a direct measurement of information loss and emphasizes the need to strike a balance between model simplicity and goodness of fit.  As a show of force, the Akaike information criterion also found the  Gutenberg-Richter law for earthquakes and the scale-free model for brain functional network, 2-dimensional sand pile, and solar flare intensity to suffer excessive loss of information. They resemble more the crumpled-together ball at low compactions in that there appear to be two driving mechanisms that take turns occurring.

\end{abstract}
\maketitle

\section{Introduction}

It is a deeply established tradition in physics to search for unifying laws, for universal principles that can bypass the specificity of particular systems to capture the underlying unity of the world. A contemporary pursuit concerns the abundant simple power-law (SPL) distributions\cite{powerlaw}, $g(x)=\alpha/ x^\beta$, over a wide range of magnitudes that surfaced in $1/f$ noise\cite{noise,abbot}, economy\cite{stanley}, distribution of income and wealth among the population\cite{pareto}, foraging patterns of sharks and tuna\cite{tuna}, and brain activity and heart rate\cite{heart} to name just a few. One attempt to explain their deeper origin is the concept of self-organized criticality  proposed by Bak, Tang, and Wiesenfeld\cite{soc} in 1987. The power-law distribution of sand avalanches and the fact that sand piles can come back to the critical slope without deliberate tuning of parameters have been a paradigm for self-organized criticality, although the dynamics of a real sand pile has been demonstrated\cite{nagel} to behave more like a first-order transition. Another notable approach is the use of renormalization group\cite{sethna}, motivated by the resemblance to the power-law divergence of physical quantities, such as specific heat, susceptibility, and correlation length,  with universal critical exponents in systems undergoing a smooth phase transition. In spite of many more generative models for various reported power laws\cite{newman}, statistical support and mechanistic sophistication are in dire need for improvement\cite{apply}. Faced with these deficiencies, it is therefore not surprising that the relevance and usefulness\cite{fox} and legitimacy\cite{abbot} of some power-law claims have been called into question.

Crackling noise from candy wrappers and food bags is something we all hate in the cinema. Its occurrence rate versus pulse intensity has also been reported\cite{kramer,houle} to obey the power law and may have bearing\cite{earthquake} on the Gutenberg-Richter law\cite{gutenberg} for earthquakes. In Section II, we shall introduce two versions of crumpling experiments. In the first one the sound data are collected from two separately crumpled thin sheets. Since the power-law exponents are distinct for different materials, we are sure that the combined data should be fit by double power laws (DPL), $\alpha_1/x^{\beta_1} +\alpha_2/x^{\beta_2}$. However, when prepared in a log-log plot,  the data points turned out to line up in an approximately straight line, and all our colleagues congratulated us for having discovered a new power law. This incident alerted us to search for a more rigorous criterion for power law and in the mean time reexamine the existing examples. Pedagogical derivations are given in Section III to explain in mathematics why the combination of two different power laws should look so tantalizingly similar to a simple power law. This paves the way for the introduction of more rigorous statistical procedures in Section IV that is capable of picking out the better fitting function among a group of competing candidates. In our second experiment two different sheets are truly crumpled together. This was briefly introduced in Section II, but can now be fully investigated after being equipped with full knowledge of the new information criterion. A change of statistical property is expected when the interactions between these two sheets increase. Initially they exhibit different power-law exponents, but as crumpling proceeds the crumpled ball should reach a compact state that is indistinguishable from that of a single (composite) sheet. In other words, we anticipate a transition of macroscopic behavior from double to a single power law as a consequence of intensifying interactions. Conclusion and discussion are arranged in Section V. Alternative and mathematically  more rigorous derivations for the results in Section III are added in Appendix A. Reasoning behind the likelihood ratio test employed to make sure the statistical evidence is strong enough to dismiss the existing model  is discussed in Appendix B. While the main text addresses only probability density functions, the statistical method we introduced in Section IV can be generalized to discuss phenomena that do not involve probability or when the raw data are not available. The procedures are detailed in Appendix C.

\section{crumple sound experiment and theoretical model}

We performed the crumple sound experiment inside a soundproof chamber with foam rubber plank on the interior to avoid echo, inside of which a microphone was connected to a Sony ICD-PX333 recorder. crumple sound is recorded at a sample rate of 44,100 points per second in 16-bit precision. The amplitude is measured in computer unit (c.u.) and maximum amplitude ($A_{\rm max}$) is $2^{15}-1$. The gain of the sound card is constant and the sample is crumpled manually at a distance of 10 cm from the microphone. We used the aluminum foil (Al), High Density Polyethylene (HDPE), and A4 copy paper as our samples. They are of thickness 16, 13, and 60$\mu$m respectively, and cut into squares (20cm$\times$20cm) for uniformity. Crumpling is kept at a slow rate of about 90 seconds per sheet to facilitate the separation of individual pulses. Care is taken to avoid friction noise caused by the relative motion between hand and sample. 

The MATLAB program was used to convert sound to signal amplitude. To estimate the background noise and dc offset, we started the recording 5 seconds prior to each round of crumpling. The average amplitude is about $10^{-3}$ as normalized by $A_{\rm max}$, and thrice this amount was set as the noise threshold. The C code algorithm automatically integrates the sound intensity every 200/44100 second. When the value exceeded that of the background noise, the beginning of a new pulse was marked. Whenever a dilemma arose at distinguishing a long pulse from two overlapping pulses, a more scrupulous criterion was applied. We resorted to smaller time step to examine the grey area by including just six amplitude peaks. Since this value is expected to decrease as a pulse fades, a sudden switch to an increasing function indicates the beginning of a second pulse.

We agree with Houle and Sethna\cite{houle} that crumple sound emits when facets suddenly buckle from one configuration to another and is not necessarily accompanied by the creation of a new ridge. To distinguish these sound-generating surfaces from the facets encircled by the ridges, we shall call the former as ``drums". Kramer and Lobkovsky\cite{kramer} who used paper that has been crumpled and uncrumpled thirty times have demonstrated that drums are a different entity from facets - a drum may comprise of many facets. To understand the origin of power law for its occurrence rate, let us imagine a sheet of unit area and call this size-1 drum. In the process of crumpling, drums of smaller sizes will appear and have their chance to emit sound at random time. Overall, we have $2^{n}$ number of size-$1/2^{n}$ drums where $n$ is a nonnegative integer. Presumably, bigger drums sound louder and we can assume the intensity of crumple sound to be proportional to the drum area; namely, $E_{n}\sim1/2^{n}$. For the sake of simplicity, we restrict each drum to  emit sound only once. The net number of sound pulses measured in the crumpling experiment consequently equals the total number of drums:
\begin{displaymath}
\sum_{n}{2^{n}}\sim\int{2^{n}dn}=\int{2^{n}\frac{dn}{dE_{n}}dE}\sim\int{\frac{1}{E^{2}}dE}
\end{displaymath}
This simple model\cite{ming} readily predicts a power law with $\beta$=2 for the occurrence rate versus sound intensity. By allowing some drums to go mute or be beaten multiple times, the exponent can be tailored to match the empirical values.

\begin{figure}
  \centering
  \includegraphics[width=8cm]{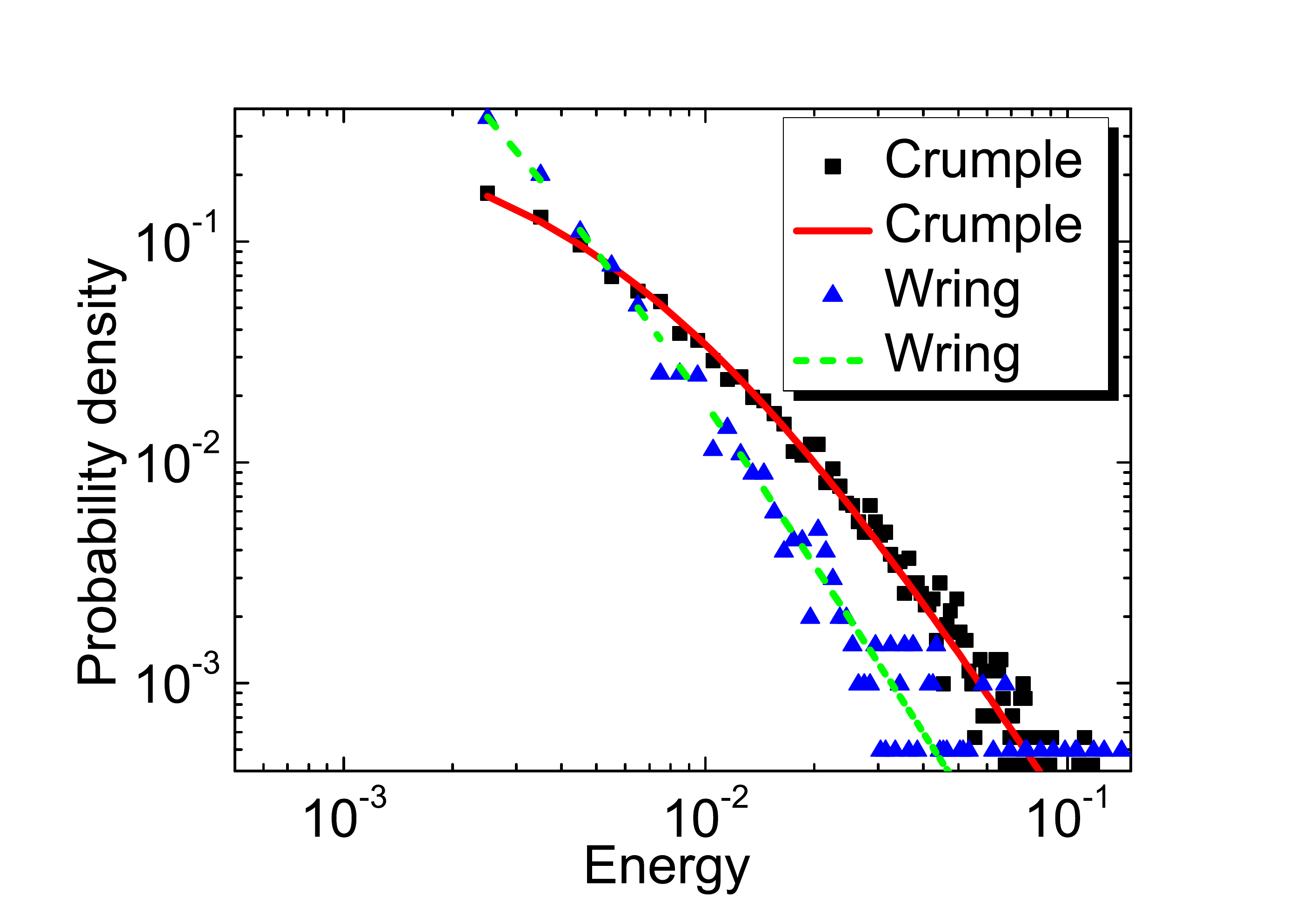} 
\caption{(color online) The normalized occurrence frequency is plotted against pulse energy for the crackling sound from a singly crumpled and wringed HDPE sheet. The red solid fitting line represents a shifted power law, while the green dotted line a simple power law. Note that both lines share the same slope at large energies, which implies their exponents are identical at about 2.653$\pm$0.006. See Table \ref{table:crumple} for further detail on the other two samples.}
\label{HDPE}
\end{figure}

\begin{figure}
  \centering
  \includegraphics[width=8cm]{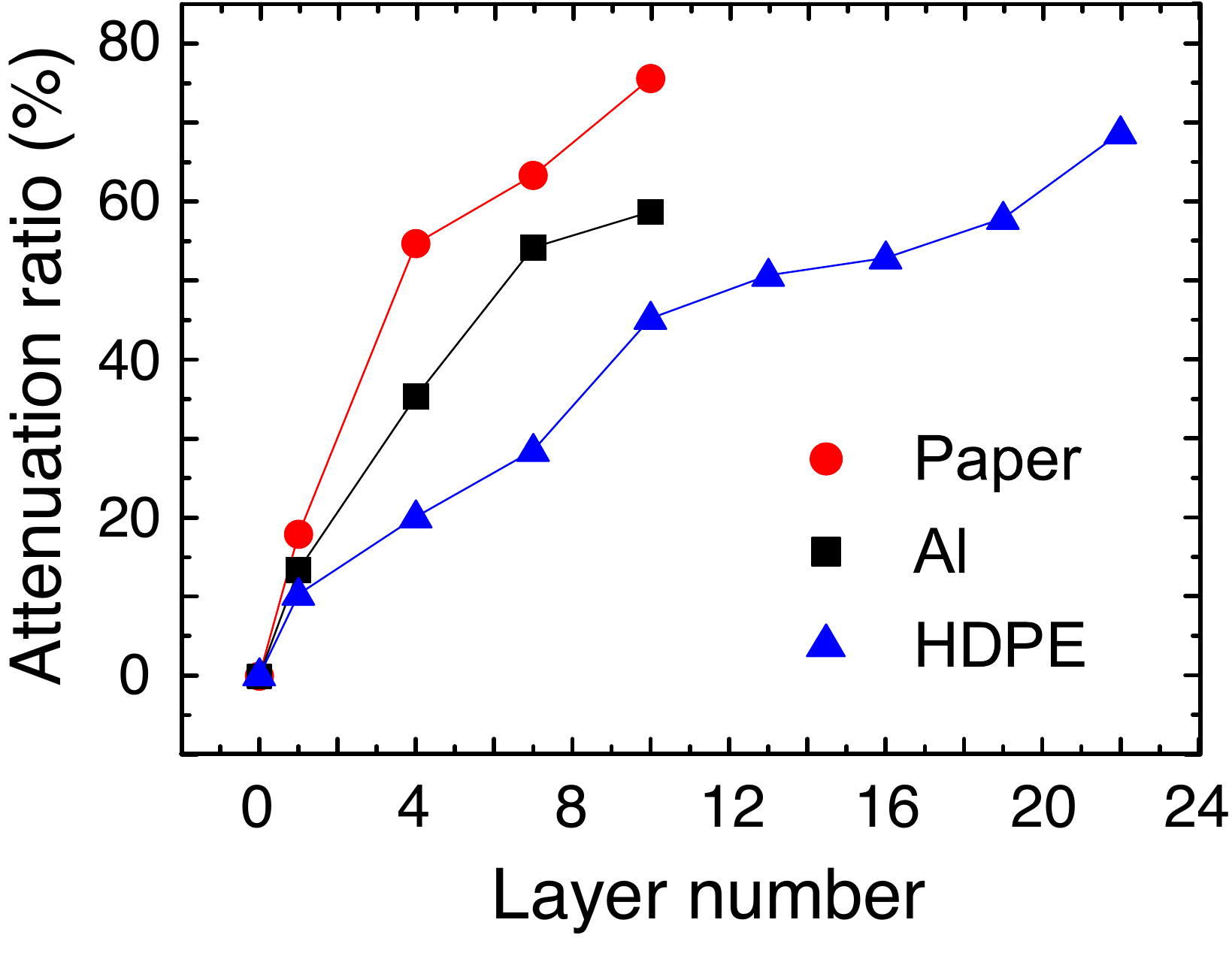} 
\caption{(color online) The ability of different material to shield sound as more layers are accumulated. A mini-speaker was inserted inside the crumpled balls to determine the attenuation ratio, as defined in the text.}
\label{atten}
\end{figure}

Unlike Ref.\cite{houle}, we found the crumple sound to differ from that from wringing via the cylinder geometry (see Fig.\ref{HDPE}). A power law can fit the wring sound nicely, but the crumple data exhibit an obvious down turn in the full-log plot and resemble more a shifted power law. We believe the discrepancy is due to the fact that a crumpled ball contains multiple layers that shield and cut down the sound intensity. Placing one or both hands over the mouth is enough to convince oneself that the attenuation by shielding must be a sizable factor. In order to quantify the effect of attenuation, we wrap the thin sheet around a mini-speaker before crumpling into  different compactions. Extent of wrinkling is therefore not fixed, but increases with the number of layers. We define the attenuation ratio as the deviation from unity of the ratio between muffled intensity and that of our prerecorded sound. The results are shown in Fig.\ref{atten}, which list paper as being most effective at dissipating the acoustic energy among the three materials for the same number of layers.

Due to the shielding effect, the straight line typical of power law in a full-log plot thus bends downward for the crumple data because  weak sounds are mostly measured in the later stage of crumpling. This is when the layer number reaches its maximum and so we expect the data points to deviate from the straight line more than their loud counterpart. This changed the distribution function from power law to a shifted power law or the Zipf-Mandelbrot distribution\cite{zipf} (ZMD), $g(x)=\alpha/(x+\gamma)^{\beta}$, where $\gamma$ is a parameter that measures the extent of attenuation. However, since ZMD reduces to power law as $x$ gets large, the $\beta$ value determined from  ZMD will be identical to that for the ``true''  model, namely, power law, without the artifact and correction due to shielding. This is verified by Fig.\ref{HDPE}.

\begin{figure}[!h]
  \centering
  \includegraphics[width=8cm]{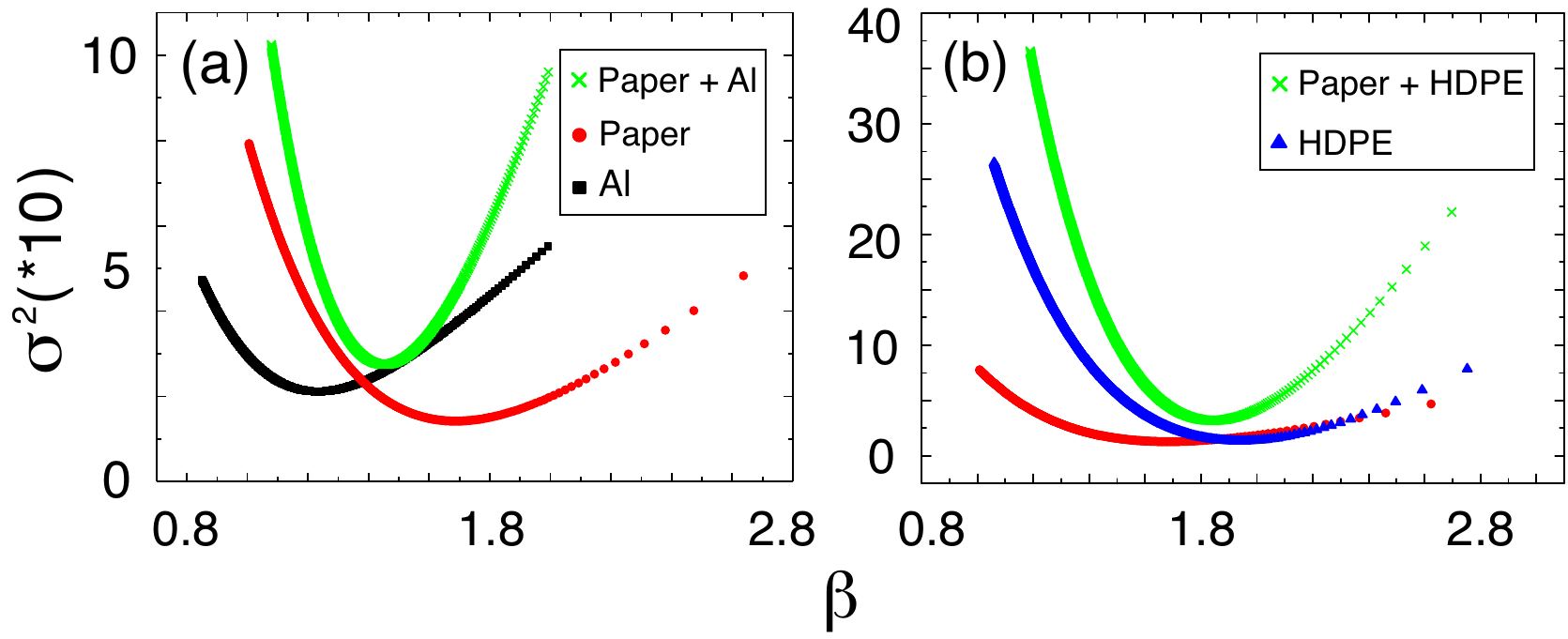} 
  \caption
  {(color online) Standard deviation plotted against fitting exponent of power law for crumple sound. $(\Delta \beta)^2$ is proportional to the ratio of   $\sigma^2 (\beta )$ and its curvature. (a) Paper (in red circle) and Al (in black square) and their combined data (in green cross). The optimal $\beta$ can be read off from the minima position to be 1.68, 1.22, and 1.44. Combined data exhibit a large curvature, which results in a misleadingly small   $\Delta \beta=0.0126$, compared to 0.0198 and 0.0200 for paper and Al. (b) Paper and HDPE (in blue triangle,  $\beta=1.94$, $\Delta \beta=0.0143$) and their combined data ($\beta=1.84$,   $\Delta \beta=0.0128$).
}\label{mixRSS} 
\end{figure}

Having established the protocol for analyzing the sound data, we move on to truly crumple two different sheets together. Although the value of $\beta$ has been determined in Fig.\ref{mixRSS} to vary with material, there is no knowing how the twisting and mingling with another sheet will affect the statistical behavior. This simple experiment is designed to simulate a complex system with two driving mechanisms and allow confirmation of the general belief and model prediction\cite{porter} that interactions can bring out a subtle change in macroscopic behavior.

\begin{figure}
  \centering
  \includegraphics[width=8cm]{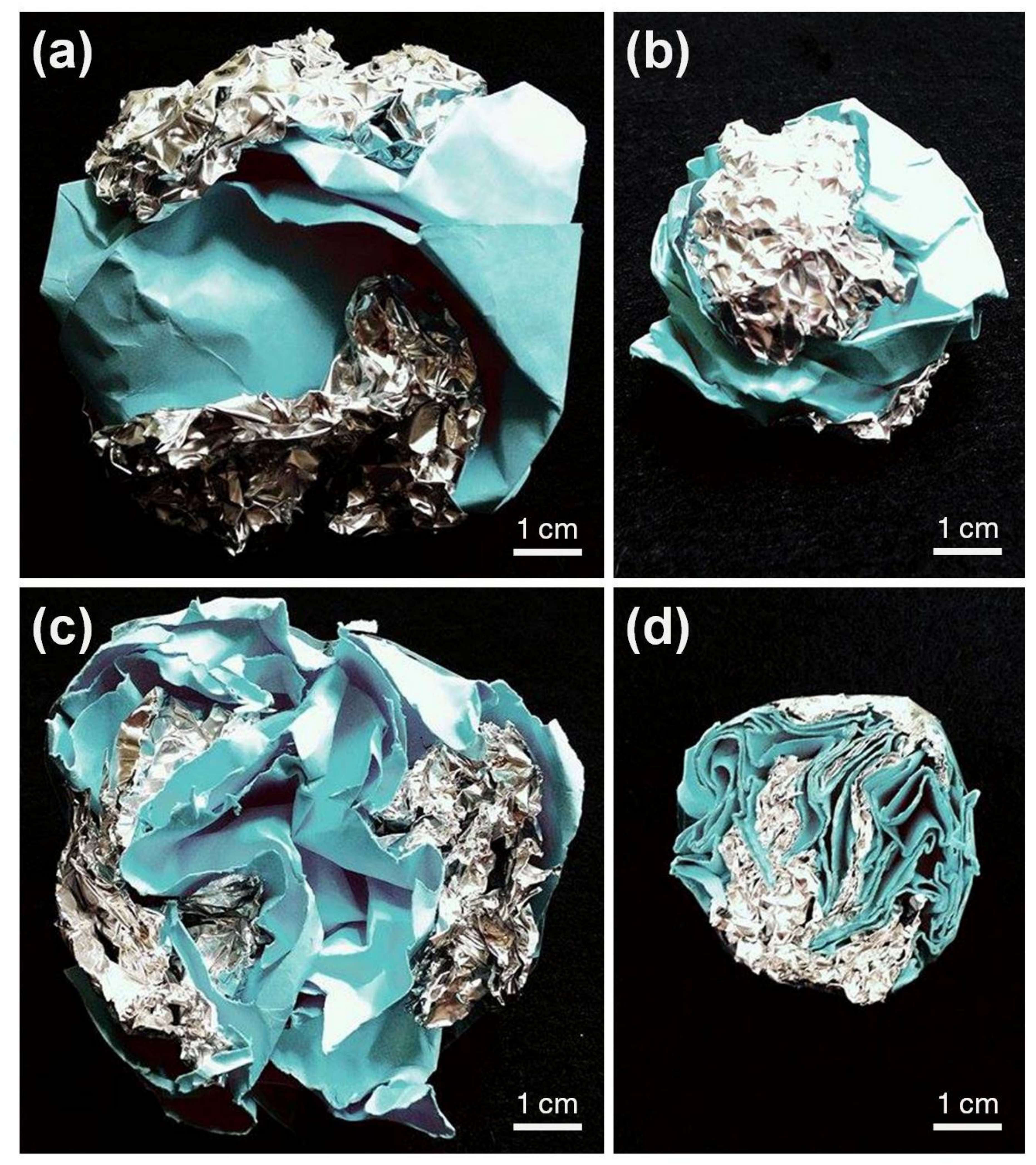} 
\caption{(color online) Aluminum foil and A4 paper (colored to enhance the contrast) are crumpled together by hand. Figures (a) and (b) show the exterior of the crumpled ball at low and high compactions, while (c) and (d) show their interior. The different characteristics of Al and paper are retained in (a) and (c) for which the occurrence rate of crumple sound is found to obey double power laws.  When inter-sheet interactions intensify, individual properties get erased and the system transits to obeying power law in (b) and (d). }
\label{cover}
\end{figure}

 In order to allow the two sheets to fully interact with each other, crumpling is thus more ideal than wringing. We oriented the two sheets in perpendicular directions prior to crumpling to prevent them from ``sticking" and becoming a single composite sheet from the beginning. Care was also taken to avoid phase separation; i.e., we made sure that both sheets mixed thoroughly and distributed evenly, as demonstrated by Fig.\ref{cover}. In contrast, a similar analysis was repeated for sheets without interactions; namely, individual hands crumpled them. It turned out that both data lined up tantalizingly straight in the full-log plot - a revealing sign of power law, while the contrast group ironically enjoyed the smaller error, $\Delta\beta$. 

\section{using a small $\Delta\beta$ as an indication for power law is prone to error}

In order to avoid this error, we need to go back and understand how $\Delta\beta$ was calculated. The magnitude of $\beta$ comes from maximizing the likelihood function $L$, while its error $\Delta \beta$ is estimated by $(-d^2 \ln L/d\beta^2)^{-1/2}$. By use of  Eq.(\ref{likelihood103}), this formula gives
\begin{equation}
\Delta \beta\sim \sqrt{\frac{\rm 2 \sigma^2}{N\frac{d^2{\rm \sigma^2}}{d\beta^2}}}
\label{variance}
\end{equation}
where $\sigma$ denotes the standard deviation. Now imagine two independent sets of power-law data, $y_i=1/x_i^{\beta_1}+\Delta y_i$ and $z_i=1/x_i^{\beta_2}+\Delta z_i$ where $\beta_1\ne\beta_2$ and $i=1,\cdots ,N$ labels the slices in the histogram. For simplicity, let us suppose the random numbers $\Delta y_i$ and $\Delta z_i$ render relatively small $\Delta \beta_k/\beta_k$ for $k=1,2$. According to Eq.(\ref{variance}),
\begin{equation}
(\Delta \beta_k)^2\sim  \frac{1}{N}\frac{\sum_{i=1}^N(\Delta y_i)^2}
                        {\sum_{i=1}^N {x_i^{-2\beta_k}}{(\ln x_i)^2}},\quad k=1,2
\label{beta1}
\end{equation}
in which a statistical average over the random numbers is implied.
If both data are combined and fit by a single power law out of ignorance, Eq.(\ref{variance}) gives
{\small
\begin{equation}
(\Delta\beta)^2 \sim \frac{\sum_i\left[(\alpha x_i^{-\beta}-x_i^{-\beta_1}-x_i^{-\beta_2})^2 +(\Delta y_i)^2+ (\Delta z_i)^2\right]}
{N \sum_i (x_i^{-2\beta_1}+x_i^{-2\beta_2}+2x_i^{-\beta_1-\beta_2})(\ln x_i)^2},
\label{beta}
\end{equation}
} 
\noindent under a general scenario in Appendix A.

In the case of perfect power laws, i.e., $\Delta y_i=\Delta z_i=0$ and $\Delta \beta_{1,2}=0$, Eq.(\ref{beta}) predicts $\Delta \beta >\Delta \beta_{i}, \forall i=1,2,$ as expected for our rash move. However, as long as $\Delta \beta_{i}\ne 0$ and exceeds about $2\times 10^{-4}$, the ratio of first terms in the numerator and the denominator of Eq.(\ref{beta}) becomes smaller than the subsequent ratio of second and third terms. A closer look reveals that the second and third ratios are simply  $\Delta \beta_{1}$ and $\Delta \beta_{2}$ from Eq.(\ref{beta1}).
It takes only simple arithmetic to confirm that $\Delta \beta < \max\{\Delta \beta_1,\Delta \beta_2\}$. 
The reasoning behind this counterintuitive result is due to the misuse of Eq.(\ref{variance}) for $(\Delta\beta)^2$ when the fitting model is wrong.
For this model misspecification situation, a correction for $(\Delta\beta)^2$ should be adopted using
Huber's sandwich estimation\cite{huber}, while applied users are seldom aware of this issue.

Figure \ref{mixRSS} shows  $\sigma^2 (\beta )$ for Al, paper, and HDPE and the combination of their data - all modeled by the power law. The much larger curvature for the combined data results in a smaller $\Delta\beta$  according to Eq.(\ref{variance}) in spite of a large $\sigma$. Due to their $\beta$ being  distinct, the correct fitting function is DPL rather than SPL. However, since increasing fitting parameter almost always improve the standard deviation, we cannot rely on the likelihood function alone to measure their relative fitting performance. It is thus imperative to seek other information criterion that also takes into account the principle of parsimony or model simplicity.

\section{Statistical analyses based on the Akaike information criterion}

Founded on information theory, the Akaike information criterion\cite{aic} came to our rescue. It quantifies the relative fitting performance of distribution functions, $g(x)$, for a given set of data. As described by the Kullback$-$Leibler distance\cite{kl} which measures the information loss when using $g(x)$ to approximate  the true distribution, AIC value is defined as
\begin{equation}
{\rm AIC}=2k-2\ln L
\label{aic}
\end{equation}
in which the first term penalizes the abuse of free parameters, $k$, and the likelihood function  $L$ rewards goodness of fit. Smaller AIC value indicates less information loss. This trade-off between goodness of fit and model simplicity  bears resemblance to the Helmholtz free energy, $F\equiv U-TS$  where $T$ denotes the temperature, for canonical ensembles in equilibrium statistical mechanics. In contrast to Eq.(\ref{aic}), $F$ balances the competing trends of minimizing internal energy $U$ and maximizing entropy $S$.

The likelihood function is defined as
\begin{equation}
L\equiv \prod^n_{i=1} g(x_i)
\label{likelihood0}
\end{equation}
in which $n$ is the number of raw data and $g(x)$ the fitting functions.  After the data has been grouped into a histogram of $N$ slices, the log-likelihood function becomes
\begin{equation}
\ln L\equiv n\sum^N_{j=1} f(x_j)\ln g(x_j),
\label{likelihood1}
\end{equation}
in which $nf(x_j)$ denotes the counts of $j$-th slice. Since both $f(x)$ and $g(x)$ are destined to describe  probability density functions, it is important to remember to impose the normalization: $\sum f(x_i)=\int g(x)dx=1$.

\begin{figure}[!h]
  \centering
  \includegraphics[width=8cm]{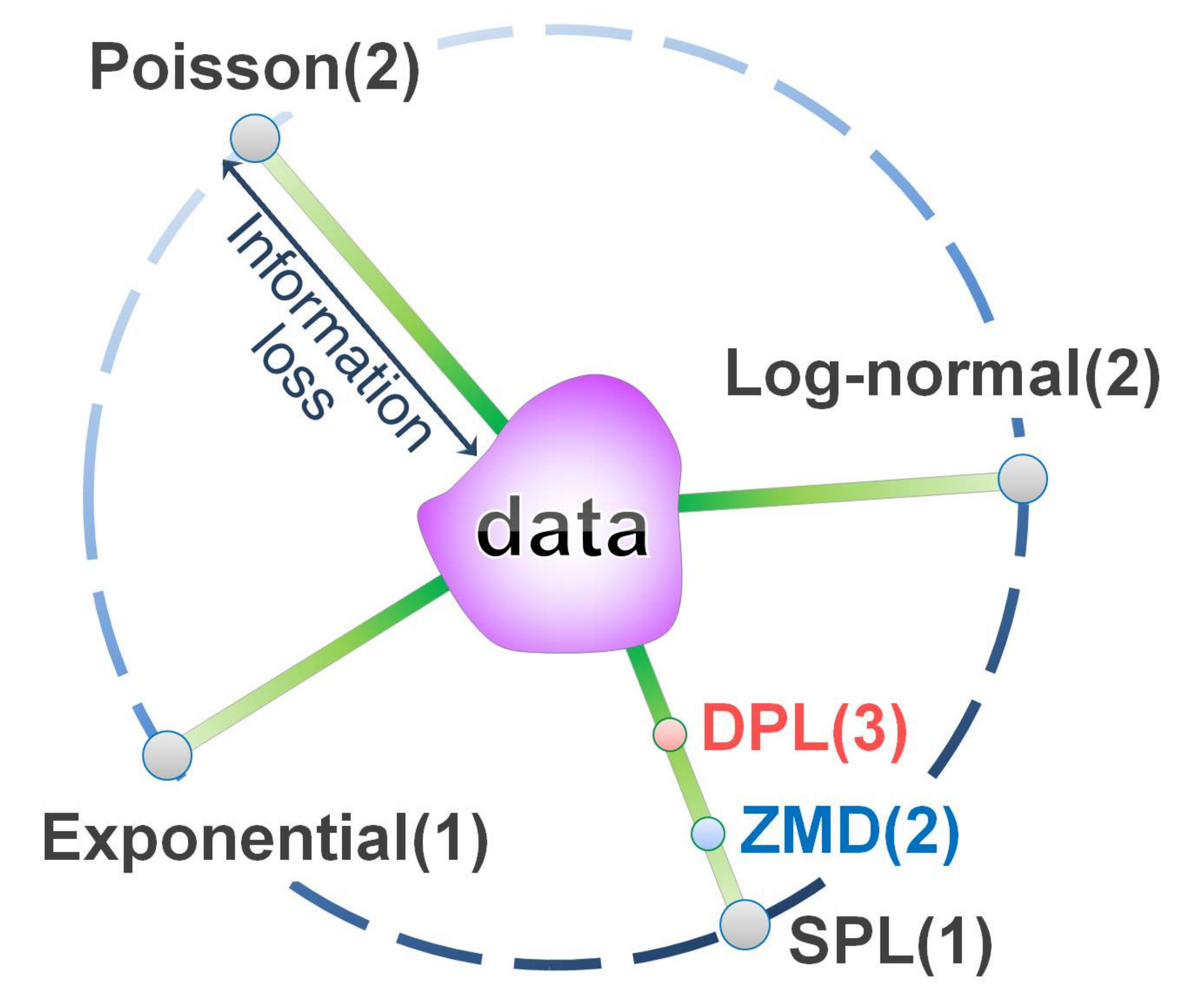} 
  \caption
  {(color online) Schematic relation of information loss by different models. Number of free parameters is indicated in the parenthesis following each distribution.  Distance between each point and the data in this model space reflects the amount of information loss as measured by the Kullback$-$Leibler distance. The dash line traces out a set of distributions with their simplest form. In contrast, DPL and ZMD are on the same green solid line as SPL since they belong to the same category. 
}\label{schematic} 
\end{figure}

 Figure \ref{schematic} illustrates the schematic relationship between DPL and ZMD, and other functions\cite{clauset} that are often checked against SPL. The ZMD and DPL are generalized versions of SPL and unlike the exponential, Poisson, and log-normal distributions that are the simplest form in their own category. However, being closer to the data in Fig.\ref{schematic}, DPL and ZMD enjoy less information loss, although they are more complex than SPL.

\begin{figure}[!h]
  \centering
  \includegraphics[width=8cm]{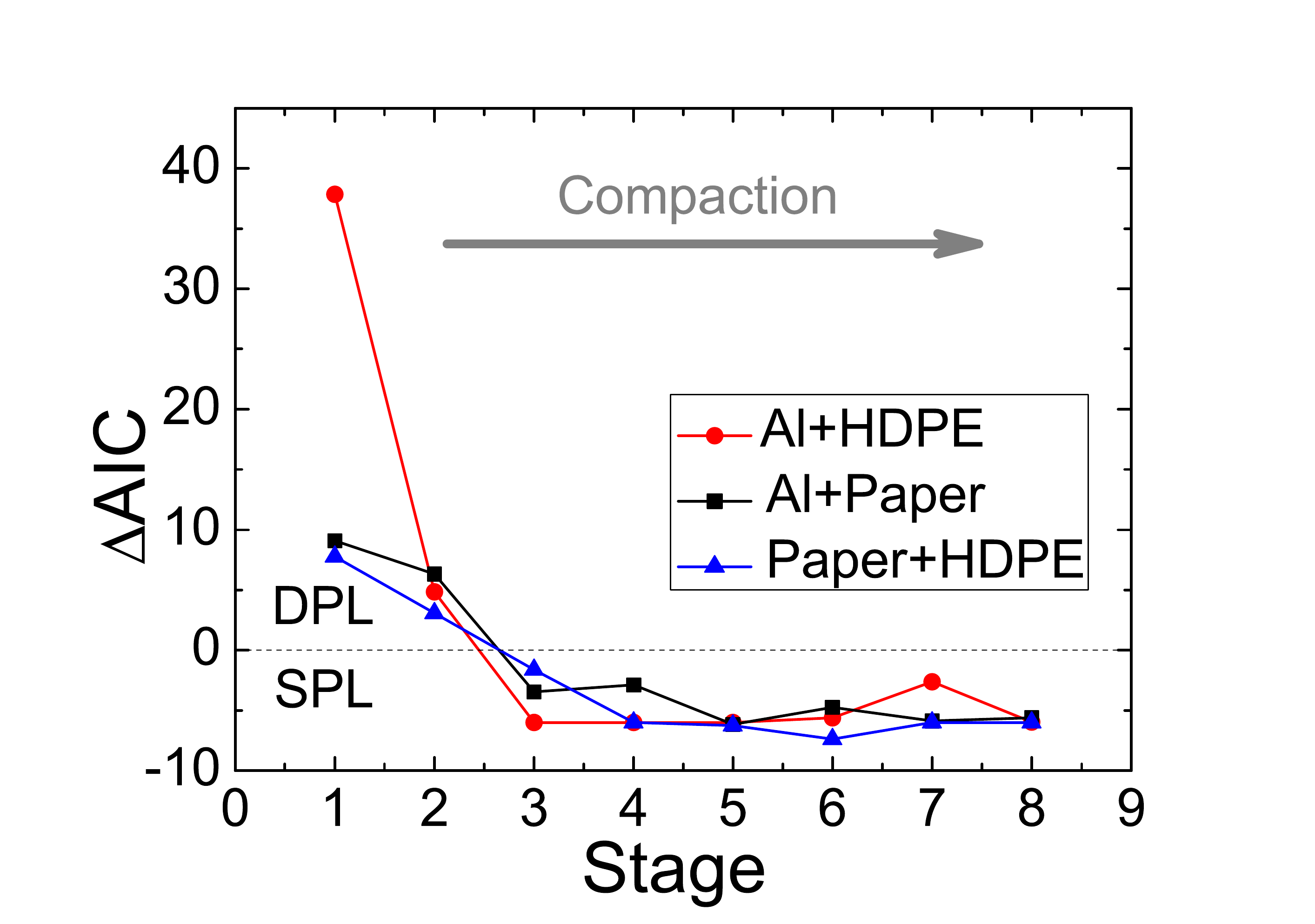} 
  \caption
  {(color online) The Akaike-information-criterion analyses of sound for two materials crumpled together. Data were divided into eight time stages with 1 being the earliest. The {\it y}-axis showed the difference of AIC value between single and double power laws. As compaction and inter-sheet interactions increases, the macroscopic behavior of the crumpled ball transits from favoring the latter to the former distribution. 
}\label{Fig5} 
\end{figure}

\begin{table*}[ht]
\caption{Comparing AIC values of different models for singly crumpled sheets. The parameters of selected model are highlighted in boldface.} 
\centering
\begin{tabular}{|  c|   c  c |  c  c| c  c |}
\hline\hline
 & \multicolumn{2}{c|}{\bf SPL} & \multicolumn{2}{c|}{\bf DPL} & \multicolumn{2}{c|}{\bf ZMD} \\ \cline{2-7}
 \bf{Sample} & \small{AIC} &  \small{$\beta$} &  \small{AIC} &  \small{ ($\beta_1$, $\beta_2$)} & \small{AIC} &  \small{($\gamma$, $\beta$)}\\ \hline
HDPE &  17379.2 & 1.732 & 16913.9 & (2.373, 2.331) & \textbf{16802.8} & \textbf{(0.0068, 2.653)} \\ \hline
Paper & 11807.8 & 1.565 & 11769.4 & (2.051, 1.894) & \textbf{11765.9} & \textbf{(0.0020, 1.797)} \\ \hline
Al & 14383.9 & 1.298 & 14376.8 & (2.249, 1.362) & \textbf{14374.4} & \textbf{(0.001, 1.369)} \\ \hline
\end{tabular}
 \label{table:crumple}
 \end{table*}

\begin{table*}[ht]
\caption{Comparing AIC values of different models for celebrated power-law claims. The parameters of selected model are highlighted in boldface. The magnitudes of $\alpha_{\rm 1,2}$  are comparable, so only their signs are included for brevity.} 
\centering
\begin{tabular}{|  c|   c  c |  c c c| c  c |}
\hline\hline
 & \multicolumn{2}{c|}{\bf SPL} & \multicolumn{3}{c|}{\bf DPL} & \multicolumn{2}{c|}{\bf ZMD} \\ \cline{2-8}
 \bf{Phenomenon} & \small{AIC} &  \small{$\beta$} &  \small{AIC} &  \small{($\alpha_1$, $\alpha_2$)} &\small{ ($\beta_1$, $\beta_2$)} & \small{AIC} &  \small{($\gamma$, $\beta$)}\\ \hline
Earthquake\cite{scsn} &  133115.7 & 1.03 & \textbf{133112.9} & (+,+) & \textbf{(1.10, 0.76)} & 133117.7 & (-0.01, 1.03) \\ \hline
Brain Functional Network\cite{brain} & 11456.73 & 2.33 & \textbf{11420.49} & (+,+) & \textbf{(3.42, 1.62)} & 11422.94 & (-3.33, 1.82) \\ \hline
Solar Flare Intensity\cite{newman} & 72028.82 & 2.10 & \textbf{71431.24} & (+,+) & \textbf{(3.76, 1.75)} & 71439.20 & (-37.07, 1.74) \\ \hline
2-dimensional Sand pile Model\cite{sand} & 1266784 & 1.01 & \textbf{1266651} & ($-$,+) & \textbf{(0.77,  0.91)} & 1266656 & (0.17, 1.04) \\ \hline
Solar Flare Rate\cite{xray} & 54747.96 & 1.10 & 54561.22 & (+,$-$) & (0.87, 0.77) & \textbf{54201.22} & \textbf{(48.08, 2.10)} \\ \hline
Web Link\cite{web} & 1095357310 & 1.72 & 1094847304 & (+,+) & (1.46, 1.71) &\textbf{1087235440} & \textbf{(0.75, 2.03)} \\ \hline
Protein-Domain Frequency\cite{zipf} & 17445.29 & 0.56 & 17427.53 &  (+,$-$) & (0.85, 0.94) &\textbf{17422.58} & \textbf{(3.89, 0.81)} \\ \hline
Stock-Market Fluctuation\cite{nyse} & 18275.57 & 3.25 & 18258.62 &  ($-$,+) & (2.43, 3.02) &\textbf{18200.94} & \textbf{(1.70, 5.82)} \\ \hline
\end{tabular}
 \label{table:aic}
 \end{table*}

Armed with the Akaike information criterion, we can quantitatively demonstrate that ZMD indeed fits the singly crumpled data better than SPL and DPL. See Table \ref{table:crumple}. We then did a more thorough analysis on the crumpled-together data by separating them into eight time stages in Fig.\ref{Fig5}. A transition from DPL to SPL was revealed as compaction increases, which confirms the prediction by Gleeson {\it et al.}\cite{porter} that interactions can bring out a subtle change in macroscopic behavior. Note that the two sheets already mingle with each other considerably at the stage when DPL was observed, as shown by Fig.\ref{cover}(c), and their emitted sounds can be easily passed for power law, if not for the scrutiny of the Akaike information criterion. The eventual switch to SPL can be understood as being characteristic of a singly crumpled {\it composite} sheet molted by the strong inter-sheet interactions. Since the exponents of DPL determined by the likelihood function match the data for the singly crumpled, the second power law cannot be dismissed as correction to SPL due to some relevant variable\cite{jim}.

In Table \ref{table:aic}, we highlighted more power-law claims that suffer excessive loss of information. Up to ten terms have been checked not to be the winner, except in the Zipf's law for word frequency\cite{word} where quadruple power laws were found to retain the most information. the Akaike information criterion concludes that DPL should replace SPL in earthquake\cite{schorlemmer}, brain functional network,  solar flare intensity, and 2-dimensional sand pile model (exemplar of self-organized criticality). 
 
\begin{figure}[!h]
  \centering
  \includegraphics[width=8cm]{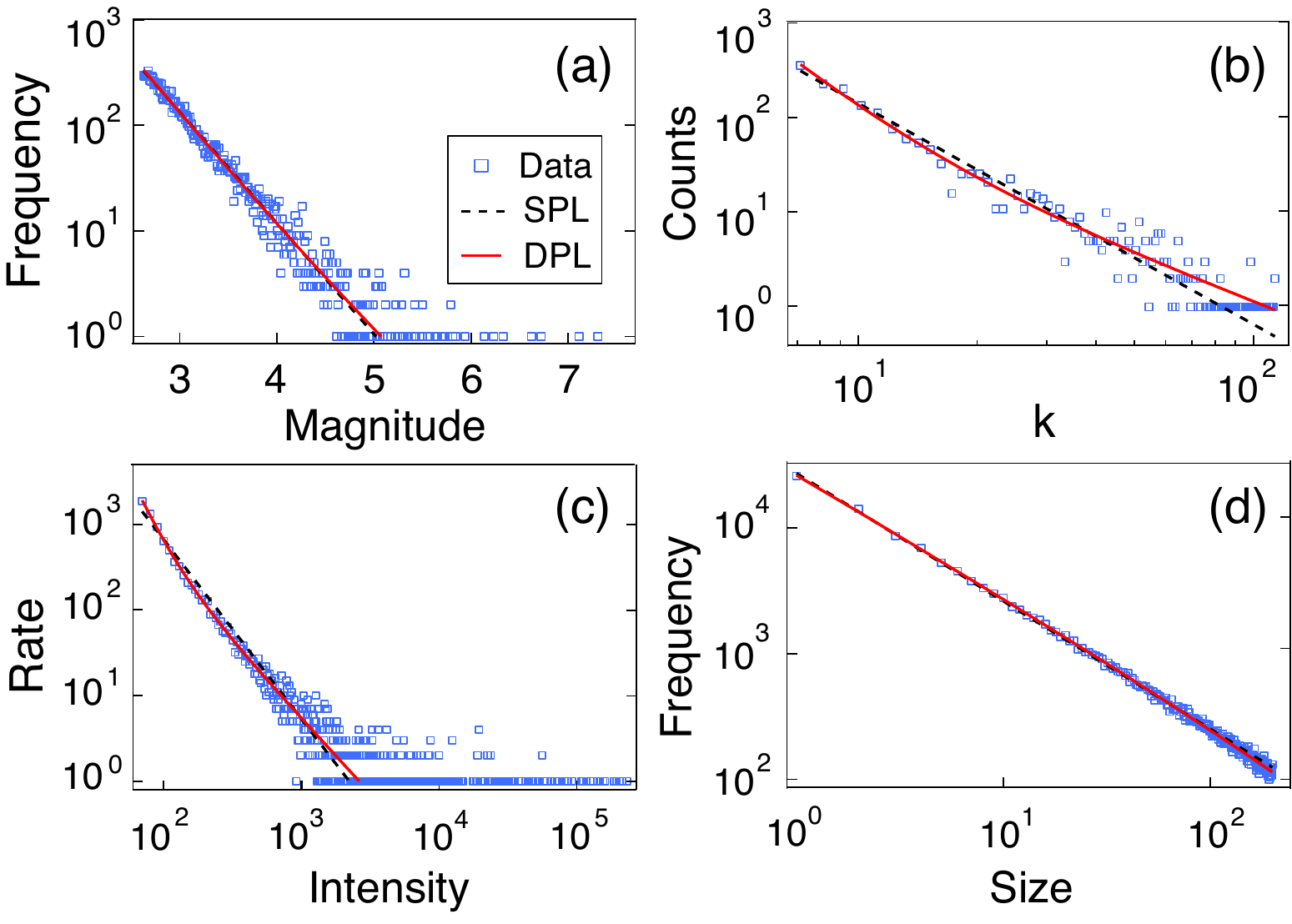} 
  \caption
  {(color online) Notable power law claims that require modifications. The DPL (in red solid line) outperforms SPL (in black dash line)  for (a) earthquakes\cite{scsn}, (b) brain functional network\cite{brain}, and (c) solar flare intensity\cite{newman}, and (d) 2-dimensional sand pile model\cite{sand}. 
}\label{cocrumple} 
\end{figure}

For the data in Fig.\ref{cocrumple}(a) that span 33 years until 2013, the Gutenberg-Richter law predicts a probability of 0.000260 for earthquakes\cite{scsn} of Richter scale 6 to 8 to occur. But according to DPL parameters in Table \ref{table:aic}, the forecast moves up by almost two folds to 0.000464. This discrepancy is statistically significant to warrant attentions. It should be noted that Schorlemmer {\it et al.}\cite{schorlemmer} concluded that the exponent of Gutenberg-Richter law could vary for different styles of faulting. However, our data were collected without screening to retain particular rake angles, and so it is not clear whether our finding can be ascribed to their theory.

Why should the brain activities in Fig.\ref{cocrumple}(b) prefer double power laws? It may simply be due to the fact that left and right hemispheres of human brain perform a fairly distinct set of operations. As for the sand pile model in Fig.\ref{cocrumple}(d), the implication is slightly trickier. Note from  Table \ref{table:aic}, the two power-law terms are of opposite signs. This implies an opposing mechanism that hinders and interferes with the ``normal'' process of self-organized criticality. We suspect the culprit is the interaction between multiple sand piles. The avalanch from one pile is sure to stack up at the foot of its neighboring piles and ``kill'' the avalanches that can originally occur there. Per Bak pointed out that the falloff at large cluster size$\sim 200$ for the 2-dimensional sand pile model is due to finite-size effect\cite{soc}. We thus have imposed an upper cutoff of 200 in Fig.\ref{cocrumple}(d) when calculating the AIC value. In other words, the fact that DPL still performs better than the simple power law implies either the finite-size effect already exists in smaller clusters or there is a yet-unknown mechanism in the sand pile model  besides the self-organized criticality. 

Table \ref{table:aic} also reported ZMD to be more favored than simple power law for the duration-time frequency of solar flare, web link, protein-domain frequency, and stock-market fluctuations. In fact, Eq.(3) in Ref.\cite{xray} already assumes the form of ZMD for the solar flare rate, but the authors neglect the shift based on the assumption of long waiting time. Adopting the same cut in the breakpoint as Ref.\cite{xray}, the Akaike information criterion still unveils ZMD as its true distribution. For the stock-market fluctuation, we followed Ref.\cite{nyse} at defining normalized stock-price return, and analyzed the daily close price of New York Stock Exchange from December 31, 1965 to May 3, 2015. Data were downloaded from
 ${\rm http://finance.yahoo.com/q/hp?s=^\wedge NYA+Historical}$ +Prices. We checked that the cumulative plot and magnitude of power-law exponent (if fit by SPL) were similar to those on pages 46 and 47 of the second work in Ref.\cite{nyse}.

Because the Akaike information criterion contains the likelihood function, it is also sensitive to the range of parameter. And the larger the range, the more effective the Akaike information criterion is at discerning the performance of different fitting functions. In order to vindicate the power-law claims in Table \ref{table:aic}, we chose the same range as their respective references.  Note that the difference in AIC values we obtained is large enough to guarantee the more stringent likelihood ratio test of which the statistical reasoning is detailed in Appendix B. The test is to make sure that the evidence is strong enough to dismiss the existing model (e.g., the SPL in this study) by requiring the following condition be met when the sample size is large\cite{sung}:
\begin{equation}
-2\ln \frac{L_1}{L_2}>\chi^2_{0.95} (k_2 -1)
\label{ratio1}
\end{equation}
where $L_1$ is the likelihood function for SPL and $\chi^2_{0.95} (k_2 -1)$ equals 5.99 (3.84) when $L_2$ represents DPL (ZMD) which corresponds to $k_2= 3$ (2).  The condition, Eq.(\ref{ratio1}), can be equivalently verified via 
\begin{equation}
{\rm AIC}_1-{\rm AIC}_2>\chi^2_{0.95} (k_2 -1)-2(k_2-1)
\label{ratio2}
\end{equation}
which clearly holds for all cases in Table  \ref{table:aic}.  For example, in order for DPL to replace SPL at describing the earthquakes, the required AIC difference is 1.99 according to Eq.(\ref{ratio2}). The value obtained in Table \ref{table:aic} is 2.8, which is large enough to justify the rejection of SPL as a null hypothesis. In the mean time, the minimum difference is 1.84 for ZMD to replace SPL.

\section{Conclusion and discussion}

We introduced the Akaike information criterion as a rigorous statistical gauge to determine whether a power law claim is legitimate. Combined with the stringent likelihood ratio test, our statistical procedures determined that several famous power laws should be rejected and replaced by double power laws, the same distribution as for the sound emission of a loose crumpled-together ball. Since we designed the latter experiment, the physical reason why the power-law ansatz is inappropriate is clear and can be explained. But, since we can not be an expert of all the phenomena in the former case that cover fields as diverse as seismology, neuroscience, astronomy, social science, and financial market, our conclusions are solely based on statistical evidence - the information theory to be specific.  In addition, the crumpled-together data are sorted temporally to reveal a transition in the statistical behavior from favoring double power laws when the inter-sheet interactions were weak to a simple power law at high compactions. This observation confirms the generally belief and  the theoretical predictions of Ref\cite{porter} that interactions can bring about a subtle change in macroscopic behavior. 

We have to admit that we do not fully understand the physical implications of our findings; e.g., what essential physics is missing in the previous models that predict the power-law behavior, why there are so few examples of thrice or higher power laws in our study, why the 2-dimensional sand pile model should betray its role as a paradigm for self-organized criticality and, more generally, how many more power laws in complex systems are in fact wrongly identified?

Derived from the Kullback-Leibler discrepancy, the Akaike information criterion provides a simple and effective way to select the best approximation model among competitors. With the optimal property of being asymptotic efficiency, it outperforms other selection criteria, such as Bayesian Information Criterion, for selecting predictive models\cite{comparison}. Based on the information theory, the Akaike information criterion is relevant to the Landauer's principle\cite{landauer} and recent interest in using entropy transfer\cite{prokopenko} to quantify directed statistical coherence between spatiotemporal processes. It can be said that the Akaike information criterion is essentially an application of the Second Law of Thermodynamics\cite{aic,burnham}, while Landauer's principle is a simple logical consequence of the law. Although the Akaike information criterion is originally intended for probability density functions, we describe in Appendix C how it can be generalized to tackle dimensional data, such as pressure versus volume for gases. 

We acknowledge funding from the Ministry of Science and Technology in Taiwan and hospitalities of the Physics Division of National Center for Theoretical Sciences in Hsinchu. We also thank Ming-Han Chou and Hsiu-Wei Yang for assistance in the early phase of this project and Hsuan-Yi Chen, Chun-Chung Chen, Jim Sethna, and Mason A. Porter for helpful comments.

\appendix

\section{Alternative and mathematically more  rigorous derivations for the results in Sec. III}
The log-likelihood for fitting SPL
\begin{displaymath}
\ln L=-\frac{N}{2}\ln \sigma^{2}-\frac{1}{2\sigma^{2}}\sum_{i=1}^N(y_{i}-\alpha x_{i}^{-\beta})^{2}
\end{displaymath}
implies

\begin{equation}
\begin{split}
& \frac{\partial \ln L}{\partial \beta}=-\frac{\alpha}{\sigma^{2}}\sum_{i}(y_{i}-\alpha x_{i}^{-\beta})x_{i}^{-\beta}\ln x_{i} \\
& \frac{\partial \ln L}{\partial \alpha}=\frac{1}{\sigma^{2}}\sum_{i}(y_{i}-\alpha x_{i}^{-\beta})x_{i}^{-\beta}
\end{split}
\label{S1}
\end{equation}
By setting zeros for Eq.(\ref{S1}), the maximum likelihood estimates $\hat{\beta}$ and $\hat{\alpha}$ satisfies
\begin{displaymath}
\hat{\alpha}=\frac{\sum_{i}y_{i}x_{i}^{-\hat{\beta}}}{\sum_{i}x_{i}^{-2\hat{\beta}}}=\frac{\sum_{i}y_{i}x_{i}^{-\hat{\beta}}\ln x_{i}}{\sum_{i}x_{i}^{-2\hat{\beta}}\ln x_{i}},
\end{displaymath}
leading to a large-sample version:
\begin{displaymath}
\alpha=\frac{\sum_{i}m_{i}x_{i}^{-\beta}}{\sum_{i}x_{i}^{-2\beta}}=\frac{\sum_{i}m_{i}x_{i}^{-\beta}\ln x_{i}}{\sum_{i}x_{i}^{-2\beta}\ln x_{i}},
\end{displaymath}
where $m_{i}\equiv x_{i}^{-\beta_{1}}+x_{i}^{-\beta_{2}}$. To obtain $\Delta \beta$ , Eq.(1) gives
\begin{equation}
\begin{split}
(\Delta \beta)^{2} &= \left[-\frac{\partial^{2} \ln L}{\partial \beta^{2}}\right]^{-1}\\
&= \frac{\sigma^{2}}{\alpha^{2}\sum_{i}x_{i}^{-2\beta}(\ln x_{i})^{2}+\alpha\sum_{i}b_{i}x_{i}^{-\beta}(\ln x_{i})^{2}}
\end{split}
\label{S2}
\end{equation}
in which $b_{i}\equiv \alpha x_{i}^{-\beta}-m_{i}$ represents the fitting bias. When the discrepancy between two SPL is small, i.e., $|b_{i}|/m_{i}<\varepsilon$ for a small $\varepsilon$, the numerator and denominator of Eq.(\ref{S2}) can be approximated by
\begin{equation}
\begin{split}
{\rm num} &= \frac{1}{N}\left[ \sum_{i}b_{i}^{2}+\sum_{i}(\Delta y_{i})^{2}+\sum_{i}(\Delta z_{i})^{2} \right]\\
{\rm den} &= \alpha^{2}\sum_{i}x_{i}^{-2\beta}(\ln x_{i})^{2}+\alpha\sum_{i}b_{i}x_{i}^{-\beta}(\ln x_{i})^{2}\\
&= \sum_{i}(\alpha x_{i}^{-\beta}-m_{i}+m_{i})^{2}(\ln x_{i})^2+\sum_{i}b_{i}(\alpha x_{i}^{-\beta}-m_{i}+m_{i})(\ln x_{i})^{2}\\
&= \sum_{i}(b_{i}+m_{i})^2(\ln x_{i})^2+\sum_{i}b_{i}(b_{i}+m_{i})(\ln x_{i})^2\\
&= \sum_{i}m_{i}^{2}(\ln x_{i}^2)+3\sum_{i}b_{i}m_{i}(\ln x_{i})^{2}+2\sum_{i}b_{i}^{2}(\ln x_{i})^{2}\\
&\sim (1+3\varepsilon+2\varepsilon^{2})\sum_{i}m_{i}^{2}(\ln x_{i})^{2}
\end{split}
\label{S3}
\end{equation}
since
\begin{displaymath}
\begin{split}
& \left|\sum_{i}b_{i}m_{i}(\ln x_{i})^{2}\right|<\sum_{i}\left|b_{i}m_{i}\right|(\ln x_{i})^{2}<\varepsilon\sum_{i}m_{i}^{2}(\ln x_{i})^{2},\\
& \left|\sum_{i}b_{i}^{2}(\ln x_{i})^{2}\right|<\varepsilon^{2}\sum_{i}m_{i}^{2}(\ln x_{i})^{2}.
\end{split}
\end{displaymath}
Plugging Eq.(\ref{S3}) into Eq.(\ref{S2}) leads to Eq.(3).

\section{Statistical reasoning behind the likelihood ratio test of Eq.(\ref{ratio1})}
AIC criteria is commonly used to rank models, which is an estimated KL distance (measuring the information loss) calculated from the empirical data. To make model comparison, one should look at the relative difference between two AIC values associated with different models, and not taking into account the AIC values themselves. The reasoning is explained via the following toy example. Suppose we have data $\{X_1,X_2,...,X_n\}$ generated from some normal distribution $N(\mu,\sigma^2)$. One would like to verify which model is preferred, Model 1 with $\mu=1$ or Model 2 with $\mu=0$? Given the definition Eq.(\ref{aic}) in the manuscript, the AIC value for Model $i$ can be simplified as
    \begin{equation}
\begin{split}
      {\rm AIC_i} &= -2\log L_i+2k_i = \underbrace{\frac{n}{2}\log (2\pi) + n}_{\mbox{constant}} \\
&+ \underbrace{\frac{n}{2}\log\left\{n^{-1}\sum_{i=1}^n(X_i-i)^2\right\}}_{\mbox{goodness of fit}}  + \underbrace{2}_{\mbox{complexity}}\\
    \end{split}
\end{equation}
    where $i=1,2$ and $k_1=k_2=1$ in this toy example.
    This expression consists of three parts: (a) a constant term related to the sample size $n$ and the normalizing factors (such as $2\pi$) in the normal density function; (b) a key term related to model fitting measuring goodness-of-fit; (c) the degrees of freedom term indicating the model complexity. The constant term plays no role on model comparison, in particular its scale is proportional to $n$ (the amount of data), which is completely irrelevant to model comparison but could be a dominate term in the AIC value when handling large data. In contrast, the goodness-of-fit term and model complexity term are critical factors for model comparison. Among these two terms, the value of goodness-of-fit growing with $n$ will further dominate the value of model complexity in determining the model ranking for handling large data sets.  Though the descriptions given above are under a particular model setting, similar arguments are generally held for other model scenarios, including our case. Back to the results given in Tables I \& II in the manuscript, the difference of AIC is small compared to AIC themselves mainly due to large sample size in our examples. But, for the purpose of model comparison, we shall only look into the difference of AIC values between models, without concerning the magnitude of AIC with an inflated constant term.

Equation (\ref{ratio1}) is a special case of a likelihood ratio test based on asymptotical distribution. Likelihood ratio test is a standard hypothesis testing method in the statistical literature. Generally speaking, it applies to a nested scenario of statistical hypotheses: the null hypothesis $H_0: \theta\in\Theta_0$ vs. the alternative hypothesis $H_1: \theta\in\Theta$, where $\Theta_0\subset\Theta$. Let $L_1$ and $L_2$ be the maximum likelihood values under the reduced model $H_0$ and the general model $H_1$, respectively.
    Based on the asymptotic theory, $-2\log(L_1/L_2)$ has a $\chi^2$ distribution with the degrees of freedom $dim(\Theta)-dim(\Theta_0)$ under $H_0$. In our case, $H_0$ refers to SPL model; $H_1$ refers to either DPL or ZMD model. Due to $dim(\Theta_0)=1$ for SPL and $k_2\equiv dim(\Theta)=2$ for DPL (or $3$ for ZMD), $-2\log(L_1/L_2)$ would behave like a $\chi^2(k_2-1)$ distribution if SPL is the underlying true model. Consequently, observing a large value of $-2\log(L_1/L_2)$ relative to the $\chi^2(k_2-1)$ distribution indicates that SPL is not plausible for the data.

Let AIC$_1$ and $L_1$ represent the AIC and maximum likelihood value for fitting SPL model, AIC$_2$ and $L_2$ represent their counterparts for fitting DPL ($k_2=2$) or ZMD ($k_2=3$) model. By definition, we have
      \begin{equation*}
      {\rm AIC_1} = -2\log L_1+2; \quad {\rm AIC_2}=-2\log L_2+2k_2.
      \end{equation*}
      Their difference satisfies
      \begin{equation*}
\begin{split}
      {\rm AIC_1}-{\rm AIC_2}&=(-2\log L_1+2)-(-2\log L_2+2k_2)\\
&=-2\log(L_1/L_2)+2(1-k_2),\\
\end{split}
      \end{equation*}
      which implies an equivalent expression of Eq.(\ref{ratio2}) from the inequality in Eq.(\ref{ratio1}).

\section{Use of the Akaike information criterion for non-probability functions}
It is always preferable to work with the raw data. However, if they are not available or when the data refer to (pressure, volume) of a gas or (pressure, mass density) of a crumpled ball, Eq.(\ref{likelihood1}) cannot be used. Reasons are simple: without information of $n$ for one thing, it is meaningless to compare $\ln L$ with $2k$ in Eq.(\ref{aic}) because Eq.(\ref{likelihood1}) now carries units. Instead, we appeal to Eq.(\ref{likelihood0}) for the Akaike information criterion.

First, suppose we know a priori that the data $(x_i, y_i)$ are close to the set $(x_i, m_\theta(x_i))$ with small errors, where $m_\theta(x_i)$ contains unknown parameters, $\boldsymbol{\theta} =(\theta_1, \theta_2, \cdots)$ . We then assume the errors, $\varepsilon_i=y_i-m_\theta(x_i)$, obey  the normal distribution and use the Gaussian likelihood in Eq.(\ref{likelihood0}). The normal assumption often holds for experimental data, in particular when data come from grouping or averaging due to central limit theorem. Under this fairly general assumption, Eq.(\ref{likelihood0}) becomes
\begin{equation}\label{likelihood100}
  L=\prod^N_{i=1} \frac{1}{\sqrt{2\pi\sigma^2}}  {\rm exp} \Bigg(-\frac{\big(y_i-m_\theta (x_i)\big) ^2} {2\sigma^2}\Bigg).
\end{equation}
By setting ${dL}/{d\sigma} =0$, it is straightforward to show that the variant maximizing the likelihood is nothing but the mean squared errors: \begin{equation}
\sigma^2 \equiv  (1/N)\sum^N_{i=1} (y_i-m_\theta(x_i))^2,
\label{likelihood102}
\end{equation}
subject to $\boldsymbol{\theta}$. By using this information and taking logarithm, Eq.(\ref{likelihood100}) reduces to 
\begin{equation}
\ln L=-(N/2)\ln (\sigma^2)-N/2.
\label{likelihood103}
\end{equation}
 The constant term $N/2$ is irrelevant when comparing different AIC values. Then AIC has a succinct form
\begin{equation}\label{AIC100}
  {\rm AIC}=2k+N\ln(\hat{\sigma}^2)
\end{equation}
where $\hat{\sigma}^2$ is the minimizer of Eq.(\ref{likelihood102}) which turns out to be equivalent to the familiar method of  least square in a regression.

If the data are accompanied with known error bars $\sigma_i$, the likelihood function in Eq.(\ref{likelihood100}) should be modified as
\begin{equation}\label{likelihood101}
  L=\prod^N_{i=1}  \frac{1}{\sqrt{2\pi{\sigma}_i^2}}  {\rm exp} \Bigg( -\frac{\big( y_i-m_\theta(x_i)\big)^2}{2{\sigma}_i^2}\Bigg)
\end{equation}
with the width of the Gaussian function input from experimental observations.
Taking logarithm now leads to the form $\ln L=C-(1/2)\sum\big( y_i-m_\theta(x_i)\big)^2/\sigma^2_i$ where constant $C$ is again not important and maximizing the second term is equivalent to minimizing the weighted least squares: 
\begin{equation}
\chi^2\equiv \sum^N_{i=1}\big( y_i-m_\theta(x_i)\big)^2/\sigma^2_i,
\label{chii}
\end{equation}
subject to $\boldsymbol{\theta}$.  AIC value becomes
\begin{equation}\label{AIC101}
 { \rm AIC}=2k+\chi^2.
\end{equation}
As a close connection, the minimizer of Eq.(\ref{chii}) can also be used to perform a $\chi^2$ goodness-of-fit test.

For the scenarios this appendix is aimed for, the raw data either have been gathered into histogram or carry units. As a result, the sample size is normally limited, i.e., the bin size $N$ might not be much larger than $k$.  In these cases the Akaike information criterion tends to pick an over-fit model and needs a bias-correction\cite{burnham,hurvich} by AICc:
\begin{equation}\label{AICc}
  {\rm AICc}={\rm AIC}+\frac{2k(k+1)}{N-k-1}
\end{equation}
with the correction term that goes to zero when $N\gg k$.

\end{document}